**Kink boundary-induced polarity revealed by nonreciprocal charge transport in $FeTe_{0.7}Se_{0.3}$ superconductor**

Shengyao Li[1,†], Ya Deng[2,†], Han Yan[2], Dianyi Hu[2], Zherui Yang[1], Xueyan Wang[1], Ming Yue[5], Qiong Wu[5*], Zheng Liu[2*], X. Renshaw Wang[1,4,*]

[1]*Division of Physics and Applied Physics, School of Physical and Mathematical Sciences, Nanyang Technological University, Singapore 637371, Singapore*
[2]*School of Materials Science and Engineering, Nanyang Technological University, Singapore 639798, Singapore*
[3]*SEU-FEI Nano-Pico Center, Key Laboratory of MEMS of Ministry of Education, Collaborative Innovation Center for Micro/Nano Fabrication, Device and System, Southeast University, Nanjing 210096, China*
[4]*School of Electrical and Electronic Engineering, Nanyang Technological University, Singapore 639798, Singapore*
[5]*Faculty of Materials and Manufacturing, Key Laboratory of Advanced Functional Materials, Ministry of Education of China, Beijing University of Technology, Beijing, 100124, China*

[†] These authors contributed equally
* Emails: wuqiong0506@bjut.edu.cn; z.liu@ntu.edu.sg; renshaw@ntu.edu.sg

**ABSTRACT**. Nonreciprocal charge transport, characterized by direction-dependent electrical response, provides a powerful probe of electronic structure and a platform for rectification. However, its realization is typically governed by intrinsic crystal symmetry, restricting the range of suitable materials. Here, we demonstrate nonreciprocal charge transport in an otherwise centrosymmetric superconductor, $FeTe_{0.7}Se_{0.3}$ (FTS), through kink boundary-induced strain engineering. The stress relaxation during growth generates a cusp-like kink-band boundary with asymmetric curvature, producing a spatially nonuniform strain field and an effective in-plane polarity that breaks inversion symmetry. Pronounced nonreciprocal transport is observed under out-of-plane magnetic fields, with strong dependence on current and temperature across the superconducting transition, in contrast to the negligible response in flat nanosheets. Our results identify kink boundary-induced symmetry breaking as a general route to engineer nonreciprocal charge transport in centrosymmetric systems.

## INTRODUCTION.

Nonreciprocal charge transport has emerged as an important class of transport phenomena, characterized by an intrinsic direction-dependent electrical response encoded within the material [1]. Such behavior is of both fundamental and technological interest, enabling current and microwave rectification without external circuit elements [2–4], while also providing a sensitive probe of underlying band structures and spin textures [5–7]. Notably, this effect can be dramatically enhanced in superconducting systems, where the relevant energy scale is reduced from the Fermi energy to the superconducting gap [8,9]. This shift has stimulated extensive efforts toward energy-efficient and high-performance superconducting rectifiers [10–13].

The realization of nonreciprocal charge transport fundamentally requires broken inversion symmetry [14]. In systems where both inversion and time-reversal symmetries are preserved, Onsager reciprocity dictates that electrical resistance remains invariant under reversal of current or magnetic field, restricting transport to reciprocal response. When inversion symmetry is broken, this constraint is lifted, allowing a nonreciprocal contribution to emerge. Phenomenologically, the resistance can be expressed as [15,16]

$$R(\boldsymbol{B},\boldsymbol{I}) = R_0(1 + \beta \boldsymbol{B}^2 + \gamma \boldsymbol{B}\boldsymbol{I}),$$

where $R_0$ is the zero-field resistance, $\beta$ characterizes conventional magnetoresistance, $\gamma$ denotes the nonreciprocal coefficient. The term proportional to $\boldsymbol{BI}$ captures the directional dependence of transport with respect to both current and magnetic field.

To harness the large nonreciprocity in superconductors, prior studies have predominantly focused on intrinsically noncentrosymmetric systems, such as crystals of polar and chiral structures [17,18], as well as engineered heterostructures or superlattices [19], where inversion symmetry is explicitly broken by design. While these platforms have yielded significant progress, the symmetry is typically fixed by material choice or growth conditions, limiting tunability and leaving other classes of superconducting systems largely unexplored.

The structural flexibility of van der Waals materials offers an alternative route to symmetry engineering, in particular, strain represents a versatile means to tailor electronic properties [20,21]. Conventional approaches rely on uniaxial stretching using pre-patterned or flexible substrates [22–24], which do not necessarily break inversion symmetry. In contrast, self-assisted three-dimensional deformations such as rolled nanotubes or wrinkles can introduce spatially nonuniform strain fields that break inversion symmetry [25,26]. While chirality-induced nonreciprocal transport has been demonstrated in rolled nanotubes [27], polarity arising from structure-induced asymmetric strain remains largely unexplored, especially in superconducting systems where vortex dynamics are highly sensitive to structural inhomogeneity [28,29].

Here, we report the emergence of nonreciprocal charge transport induced by the kink boundary in synthesized $FeTe_{0.7}Se_{0.3}$ (FTS) nanosheet. High-quality FTS nanosheets are synthesized via chemical vapor deposition, where stress relaxation during growth leads to the formation of a cusp-like kink-band boundary at the center of the flake. A pronounced nonreciprocal response is observed during the superconducting transition, in stark contrast to the negligible signal in flat FTS nanosheets. Detailed analysis shows that the dominant response couples to the out-of-plane magnetic field component, consistent with an effective in-plane inversion-breaking response generated by asymmetric strain across the boundary. The observed nonreciprocity is attributed to the combined contributions of asymmetric vortex dynamics and superconducting fluctuation induced paraconductivity. These results establish strain-engineered anomalous boundaries as an effective route to induce and control nonreciprocal transport in otherwise centrosymmetric superconductors, opening new possibilities for designing superconducting rectifiers.

**RESULTS**

The iron-based superconductor FTS has attracted sustained interest owing to its relatively high critical temperature ($T_c$) and nontrivial surface states [30,31]. Recent advances in chemical vapor deposition (CVD) growth of FTS nanosheets further establish it as a promising platform for scalable superconducting device integration [32,33]. Through controlled growth conditions during CVD process, the initially two-dimensional FTS nanosheet evolves into a three-dimensional kink-boundary structure, as illustrated in Fig. 1(a). A magnified transmission electron microscopy (TEM) image reveals crystalline continuity across the boundary, confirming the structural integrity of the nanosheet despite the large deformation.

The formation of the three-dimensional kink boundary introduces spatially nonuniform strain governed by the local curvature on either side of the ridge. Atomic force microscopy (AFM) measurements capture this geometric asymmetry. As shown in Fig. 1(b), the flat region (region 2) has a thickness of 60 nm, while cusp-like kink boundary (region 1) exhibits a height of 1664 nm with markedly asymmetric slopes on the two sides. Despite its three-dimensional morphology, the kink is formed by a continuous FTS nanosheet rather than a bulk volume, therefore can be regarded as a curved, quasi-two-dimensional superconductor. The superconducting properties of the kink-boundary nanosheet are characterized via temperature-dependent resistance measurements [Fig. 1(c)], exhibiting $T_c$ = 12.5 K, defined by the midpoint of the resistive transition. To examine the role of phase-fluctuations near the superconducting transition, the temperature-dependent *V-I* characteristics are analyzed using the power-law relation $V \propto I^{\alpha(T)}$. Fig. 1(d) presents the *V-I* curves on logarithmic scales over the temperature range from 10.5 to 12.8 K. The exponent $\alpha$ was extracted by linear fitting of the power-law regime, as indicated by the black line. As shown in Fig. 1(e), $\alpha$ decreases with increasing temperature and reaches the Berezinskii-Kosterlitz-Thouless (BKT) criterion $\alpha = 3$ at $T_{\mathrm{BKT}} \approx 11.9\,\mathrm{K}$, slightly below the superconducting transition temperature $T_c \approx 12.5$ K, consistent with BKT-like phase ordering and appreciable vortex-mediated phase fluctuations near the superconducting transition.

Pristine FTS crystallizes in the tetragonal P4/nmm space group, where the Fe-Te/Se sublattice preserves inversion symmetry. According to Onsager's reciprocal relations, charge transport in such a centrosymmetric system remains invariant under the reversal of magnetic field and current direction. Figure 1(f) schematically depicts this scenario in a flat FTS nanosheet, along with the expected second-harmonic resistance ($R^{2\omega}$) response, which has been widely adopted to probe nonreciprocal charge transport [9]. Under an applied AC current of frequency ω, the second-harmonic voltage at 2ω is detected via lock-in techniques, yielding

$$R^{2\omega} = \frac{1}{2}\gamma R_0 B I_0, \quad (1)$$

where $\gamma$ denotes the nonreciprocal coefficient, $R_0$ the linear resistance, $B$ the magnetic field, and $I_0$ the current amplitude. For a centrosymmetric lattice ($\gamma = 0$), the second-harmonic response vanishes irrespective of current amplitude or magnetic field, indicating purely reciprocal transport in the flat FTS nanosheet.

Strain engineering has long been recognized as an effective route to modulate carrier mobility, electronic band structure, and magnetic properties in two-dimensional materials [21]. Of particular relevance, spatially nonuniform strain can generate a strain gradient that breaks inversion symmetry and produces an effective polarity, thereby enabling directional responses such as the flexo-photovoltaic effect [34,35], and nonreciprocal charge transport [13]. This mechanism differs from conventional uniform-strain engineering, for which a macroscopic strain gradient and the associated flexoelectric polarity are absent. Artificial structures based on step edges, local point loading, or controlled bending have therefore been developed to create nonuniform strain fields. Here, the three-dimensional kink boundary provides a natural pathway to break inversion symmetry. As schematically illustrated in Fig. 1(g), the asymmetric curvature and associated nonuniform strain distribution of the kink break inversion symmetry across the boundary, giving rise to a predominantly in-plane polarity $\boldsymbol{P}$. This emergent polarity provides the fundamental condition required for nonreciprocal charge transport in the form of $\Delta R \propto (\boldsymbol{P} \times \boldsymbol{B}) \cdot \boldsymbol{I}$. Correspondingly, the lower panel of Fig. 1(g) shows a clear linear dependence of $R^{2\omega}$ on current amplitude in the kink-boundary nanosheet, with the signal reversing sign upon magnetic field inversion.

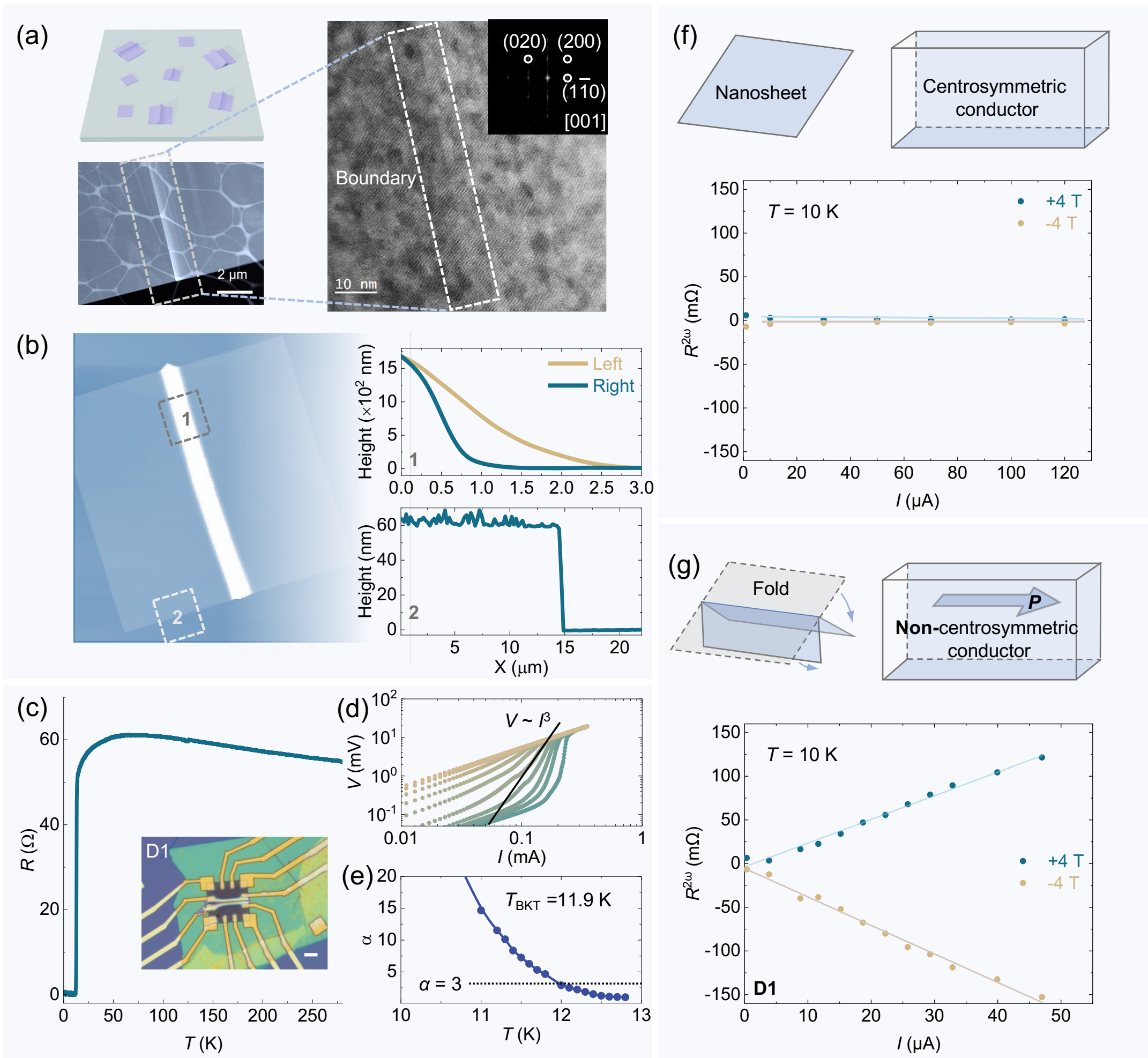

FIG. 1. Nonreciprocal charge transport in kink-boundary FTS nanosheets. (a) Schematic illustration and transmission electron microscopy (TEM) image of an FTS nanosheet transferred onto a Cu grid. The inset shows a magnified view of the boundary region, together with the corresponding fast Fourier transform (FFT) pattern, confirming single crystallinity along the [001] direction. (b) Atomic force microscopy (AFM) height profile of the nanosheet with a kink boundary, revealing a pronounced cusp with asymmetric slopes across the boundary. (c) Temperature-dependent resistance of the kink-boundary nanosheet, showing a superconducting transition at $T_c = 12.5$ K. Inset: optical image of the device (scale bar: 10 μm). (d) Temperature-dependent *V*-*I* characteristics on logarithmic scales. (e) Extracted exponent $\alpha$ as a function of temperature. The criterion $\alpha = 3$ yields $T_{\mathrm{BKT}} \approx 11.9$ K. (f) $R^{2\omega}$ measured in pristine flat FTS nanosheet, exhibiting reciprocal charge transport. (g) $R^{2\omega}$ measured across the boundary, displaying a linear dependence on current amplitude and characteristic nonreciprocal behavior.

To elucidate the role of the kink boundary, we compare three configurations: current applied across the boundary, along the boundary, and through a flat FTS nanosheet. Figures 2(a)-(c) show the temperature-dependent resistance under out-of-plane magnetic fields up to 7 T. All three configurations exhibit comparable zero-field transition temperatures of $T_c = 12.5$ K, while the transitions broaden progressively with increasing field. The resistive state develops more rapidly in the flat nanosheet than the kink-boundary devices, suggesting that the kink may suppress vortex motion and reduce the flux-flow dissipation.

Figures 2(d)-(f) compare $R^{\omega}$ and $R^{2\omega}$ as a function of out-of-plane magnetic field at $T$ = 11 K. In all configurations, $R^{\omega}$ increases with $\boldsymbol{B}$ as superconductivity is progressively suppressed. At $B$ = 7 T, $R^{\omega}$ reaches approximately 18%, 7% and 38% of the corresponding normal-state resistance for current across the boundary, along the boundary, and through the flat nanosheet, respectively. The stronger retention of the low-resistance state in kink-boundary devices is consistent with suppressed vortex motion near the boundary. Both kink-boundary configurations exhibit pronounced second-harmonic responses that are antisymmetric under magnetic-field reversal [Fig. 2(d) and 2(e)], whereas $R^{2\omega}$ remains negligible in the flat nanosheet [Fig. 2(f)].

Although the $R^{2\omega}$ signal is larger for current across the boundary than for current along it, its magnitude is also affected by the different dissipative resistances of the three configurations. We therefore evaluate the effective nonreciprocal coefficient $\gamma$. To avoid the numerical uncertainty associated with the small values of $\boldsymbol{B}$ and $R^{\omega}$ near zero field, we restrict the comparison to fields above 1 T. As shown in Figs. 2(g) and 2(h), the two kink-boundary configurations exhibit comparable $\gamma$ over most of the measured field range, despite their different $R^{\omega}$ and $R^{2\omega}$. By contrast, the flat nanosheet exhibits a much smaller $\gamma$ that is approximately two orders of magnitude below that of the kink-boundary devices. These results demonstrate that the enhanced nonreciprocal response is associated with inversion-symmetry breaking at the kink boundary rather than simply with the magnitude of the linear resistance. Meanwhile, the comparable nonreciprocal responses for current parallel and transverse to the boundary suggest that the kink boundary could enable multiple nonzero components of the nonlinear transport tensor, whose relevant magnitudes depend on the detailed boundary morphology, local strain distribution, and current density profile (See Supplemental Material [36] for additional measurements under different current and magnetic field orientations).

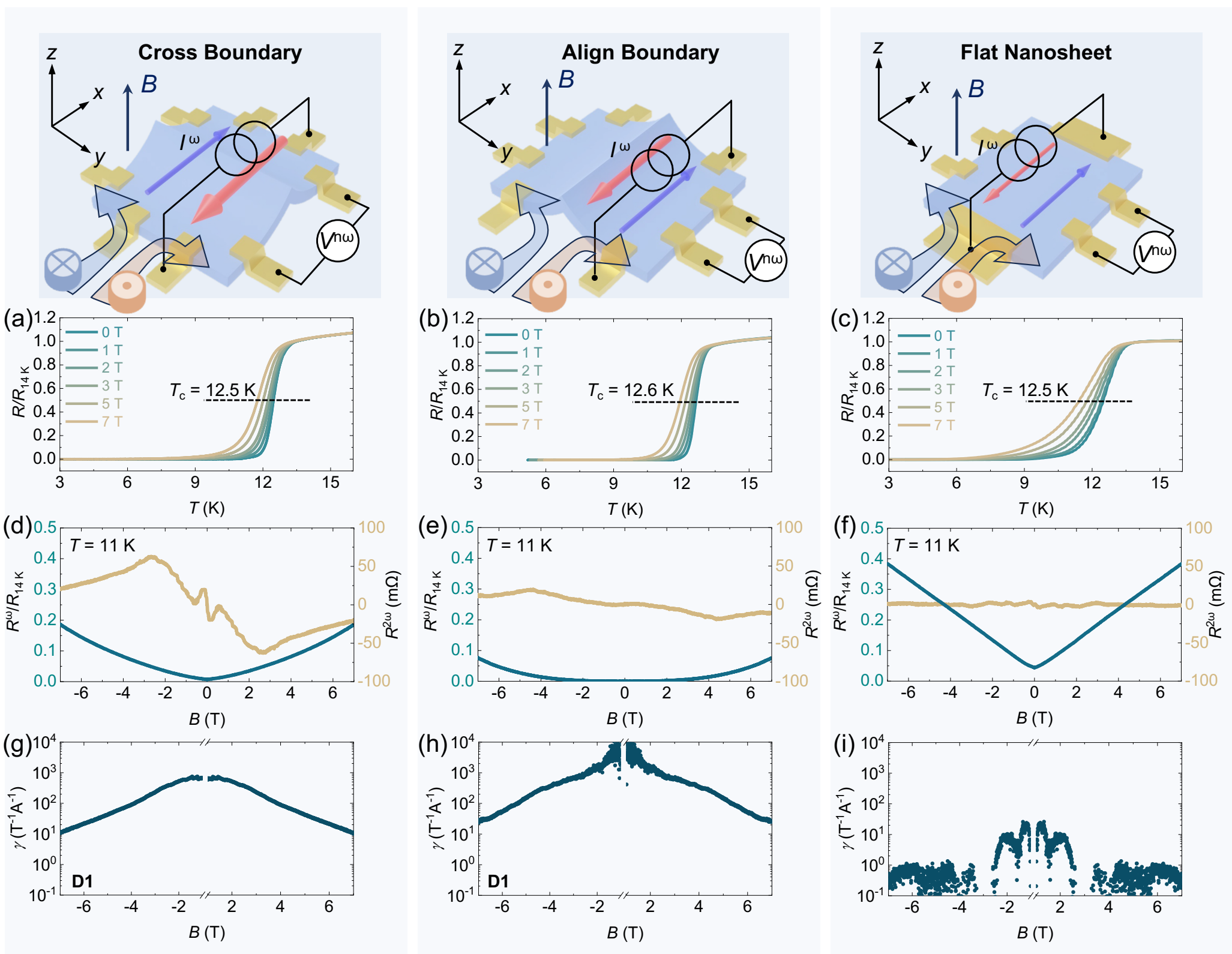

FIG. 2. Orientation-dependent magnetotransport of kink-boundary and flat FTS devices. Schematics and transport measurements for current applied across the kink boundary (left column), along the boundary

(middle column), and through a flat nanosheet (right column). (a)-(c) Temperature dependence of resistance under out-of-plane magnetic fields from 0 to 7 T. (d)-(f) Magnetic-field dependence of $R^{\omega}$ (left axes) and $R^{2\omega}$ (right axes) at $T$ = 11 K. (g)-(i) Corresponding effective nonreciprocal coefficient $\gamma$ as a function of magnetic field.

The microscopic origin of nonreciprocal charge transport is intimately tied to crystal symmetry, leading to distinct angular dependencies on magnetic field and current in chiral and polar systems. In chiral conductors, the nonreciprocal response is maximized when the relevant vectors (current and magnetic field) are collinear, whereas in polar systems the leading response commonly involves mutually orthogonal polarity, magnetic field and current components [37,38].

In the present FTS kink boundary [Fig. 3(a)], the asymmetric curvature on the two sides of the boundary is expected to generate an effective polarity $\boldsymbol{P}$. To determine the orientation of the kink-induced polarity, we performed angular-dependent measurements in the geometry shown in Fig. 3(a), where magnetic field is rotated within the polar and azimuthal planes.

Figure 3(b) shows the magnetic field-dependent $R^{2\omega}$ at $T$ = 12 K for different polar angles $\theta$. The device remains in the finite-resistance regime over the relevant field range, allowing vortex-related dissipation and the associated nonlinear responses. As the field is tilted from the out-of-plane toward the in-plane direction, the $R^{2\omega}$ peaks decrease systematically and nearly vanish for an in-plane field. Their sign reverses when $B_z$ changes sign, showing that the dominant second-harmonic response tracks the sign of the out-of-plane field component. In contrast, azimuthal rotation of the field within the sample plane produces no systematic evolution of $R^{2\omega}$ [Fig. 3(c)], and the signal remains substantially smaller than that obtained under an out-of-plane field.

The angular dependences of the $R^{\omega}$ and $R^{2\omega}$ responses are further presented in Figs. 3(d) and 3(e) as a function of magnetic-field magnitude and polar angle. Although $R^{\omega}$ evolves continuously with field and remains finite near the in-plane orientation, $R^{2\omega}$ is strongly suppressed in this angular region even at fields up to 7 T. Thus, the absence of $R^{2\omega}$ cannot be attributed simply to the absence of dissipation or to the magnitude of the linear resistance. To show the full angular dependence, Fig. 3(f) exhibits the polar angular dependence of $|R^{2\omega}|$ at $B$ = 2 T. The polar-angle dependence exhibits a pronounced two-fold anisotropy, with maxima for an out-of-plane field and nearly zero for an in-plane field.

To further account for differences in the dissipative resistance, we extract the peak value of $R^{2\omega}$ from each polar-angle field sweep and evaluate the effective nonreciprocal coefficient $\gamma$ as a function of polar angle. As shown in Fig. 3(g), $R^{2\omega}$ rises sharply upon a slight tilt 5° away from the plane, eventually saturating as the field approaches the out-of-plane direction, while $\gamma$ exhibits a similar monotonic increase with the tilt angle. These results establish that the dominant nonreciprocal response couples to $B_z$, consistent with previously reported behavior in intrinsic polar superconductors [11,39,40], supporting the interpretation that the asymmetric strain distribution across the kink boundary generates an effective in-plane polarity.

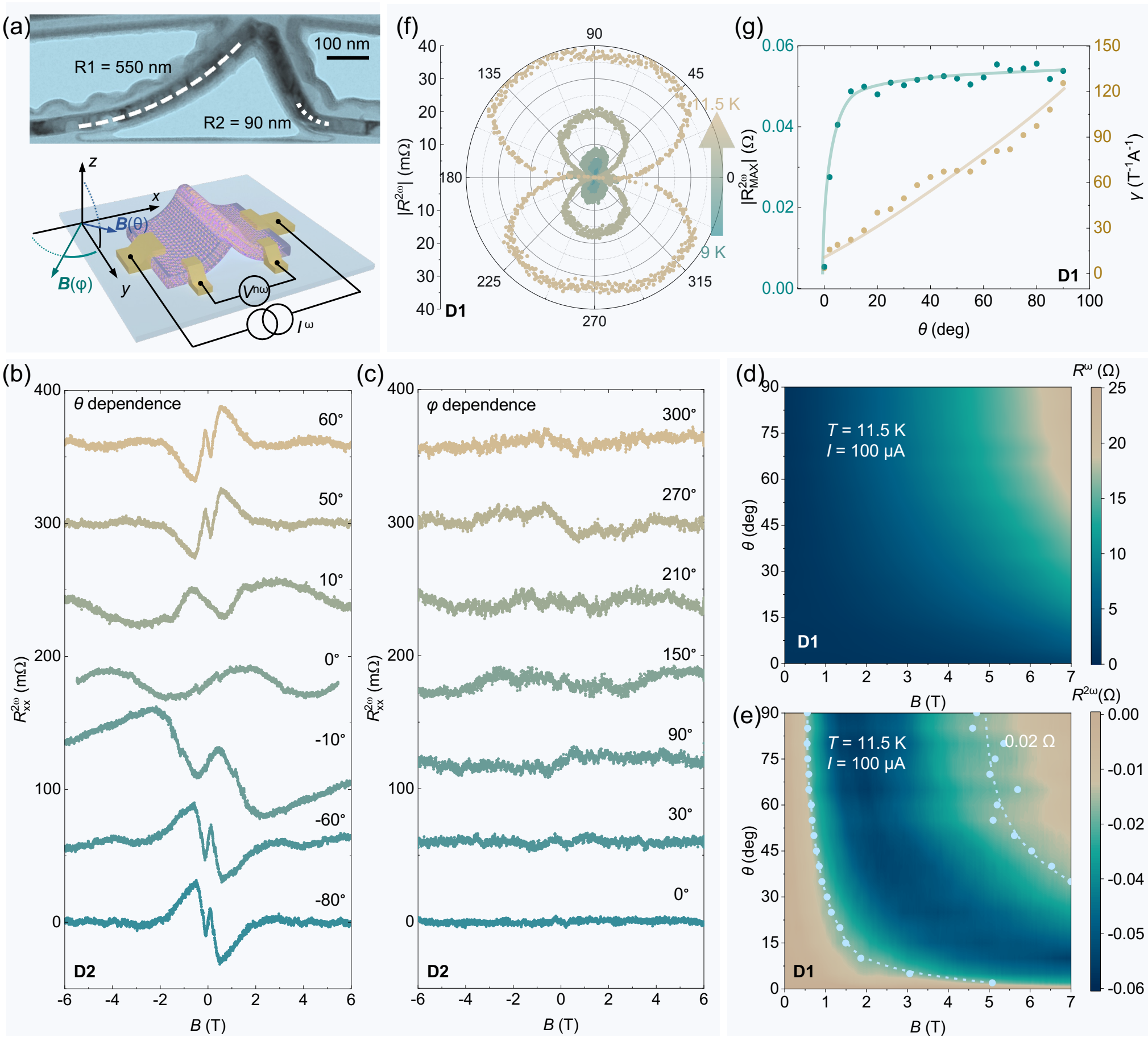

FIG. 3 Angular dependence of kink-induced nonreciprocal transport. (a) Cross-sectional image of the kink boundary and schematic of the measurement configuration for angular-dependent measurements. (b) Magnetic field dependence of $R^{2\omega}$ at different polar angles at $T$ = 12 K. (c) Magnetic field dependence of $R^{2\omega}$ at different azimuthal angles at $T$ = 12 K. (d) Contour plots of $R^{\omega}$ as functions of magnetic field magnitude and polar angle at $T$ = 11.5 K. (e) Corresponding contour plots of $R^{2\omega}$. (f) Polar plots of $|R^{2\omega}_{MAX}|$ at $B$ = 2 T at different temperatures. (g) Peak $|R^{2\omega}|$ and corresponding $\gamma$ as a function of polar angle at $T$ = 11.5 K.

Nonreciprocal charge transport in superconductors is of particular interest due to the coexistence of multiple contributing mechanisms. On the one hand, inversion-symmetry breaking leads to band splitting, which enhances the nonreciprocal response in the superconducting fluctuation regime through parity mixing and modification of the superconducting gap, commonly described within the framework of paraconductivity [41]. On the other hand, vortex dynamics in the mixed state provide an additional and often dominant contribution, giving rise to rich nonreciprocal behavior associated with vortex motion. Both mechanisms are tied to the superconducting transition and depend strongly on magnetic field and current.

Figure 4(a) presents the magnetic-field dependence of $R^{\omega}$ and $R^{2\omega}$ at temperatures ranging from 10 to 12.7 K. Over a broad temperature range, $R^{\omega}$ increases gradually with magnetic field without exceeding the normal-state resistance $R_N$, indicating robust superconductivity. In contrast, the magnitude of $R^{2\omega}$ increases with temperature, reaching a maximum around 12 K, and then rapidly diminishes to nearly zero above $T_c$. The extracted temperature-dependent nonreciprocity is shown in Fig. 4(b). The pronounced enhancement of $\gamma$ near the superconducting transition highlights the critical role of superconducting fluctuations in amplifying the nonreciprocal response.

The current dependence is examined in Fig. 4(c), which shows $R^{\omega}$ and $R^{2\omega}$ measured at $T = 10$ K under excitation currents ranging from 50 to 250 μA. Increasing the current not only drives vortex motion but also enhances nonequilibrium superconducting fluctuations, thereby modulating the paraconductivity. As the current increases, the resistive state expands, reflected by a rise in $R^{\omega}$. Correspondingly, $R^{2\omega}$ initially increases with current and then gradually saturates. The extracted peak values of $R^{2\omega}$ and $\gamma$ are summarized in Fig. 4(d). In the low-current regime (<100 μA), $R^{2\omega}$ scales approximately linearly with current. At higher currents, the response saturates and eventually decreases, which can be attributed to the weakening of vortex pinning [41,42].

To disentangle the respective contributions from vortex dynamics and paraconductivity, the temperature dependence of $R^{\omega}$, $R^{2\omega}$, and $\gamma$ is analyzed under an out-of-plane magnetic field of $B = 1.5$ T. At low temperatures, vortices are strongly pinned by defects, resulting in a zero-resistance state [43]. As temperature increases, thermal activation enables vortices to overcome pinning barriers, leading to finite dissipation and the emergence of $R^{\omega}$. In noncentrosymmetric superconductors, moving vortices experience direction-dependent energy barriers, giving rise to a nonreciprocal response commonly described as a vortex ratchet effect [44,45]. In this regime, $R^{2\omega}$ follows the evolution of $R^{\omega}$ and reaches a maximum near 12.15 K, suggesting a substantial contribution from vortex ratchet motion.

Notably, a kinked feature appears slightly below $T_c$, where superconducting amplitude fluctuation is most pronounced, suggesting an additional contribution from paraconductivity [8]. Figure 4(f) shows an enlarged view of $\gamma$ near the transition, together fitted using [39]

$$\gamma(T) = \gamma_s\left(1 - \frac{R(T)}{R_N}\right)^2, \tag{2}$$

where $R_N = 59\ \Omega$ (at $T = 14$ K), $\gamma_s$ is the saturation value of $\gamma$ at $T = 12$ K. The deviation between the fitted curve and the experimental data below 12.4 K is attributed to the additional contribution from vortex rectification. These results indicate that the observed nonreciprocal response in the kink boundary FTS nanosheet arises from the combined effects of vortex dynamics and superconducting fluctuation-induced paraconductivity.

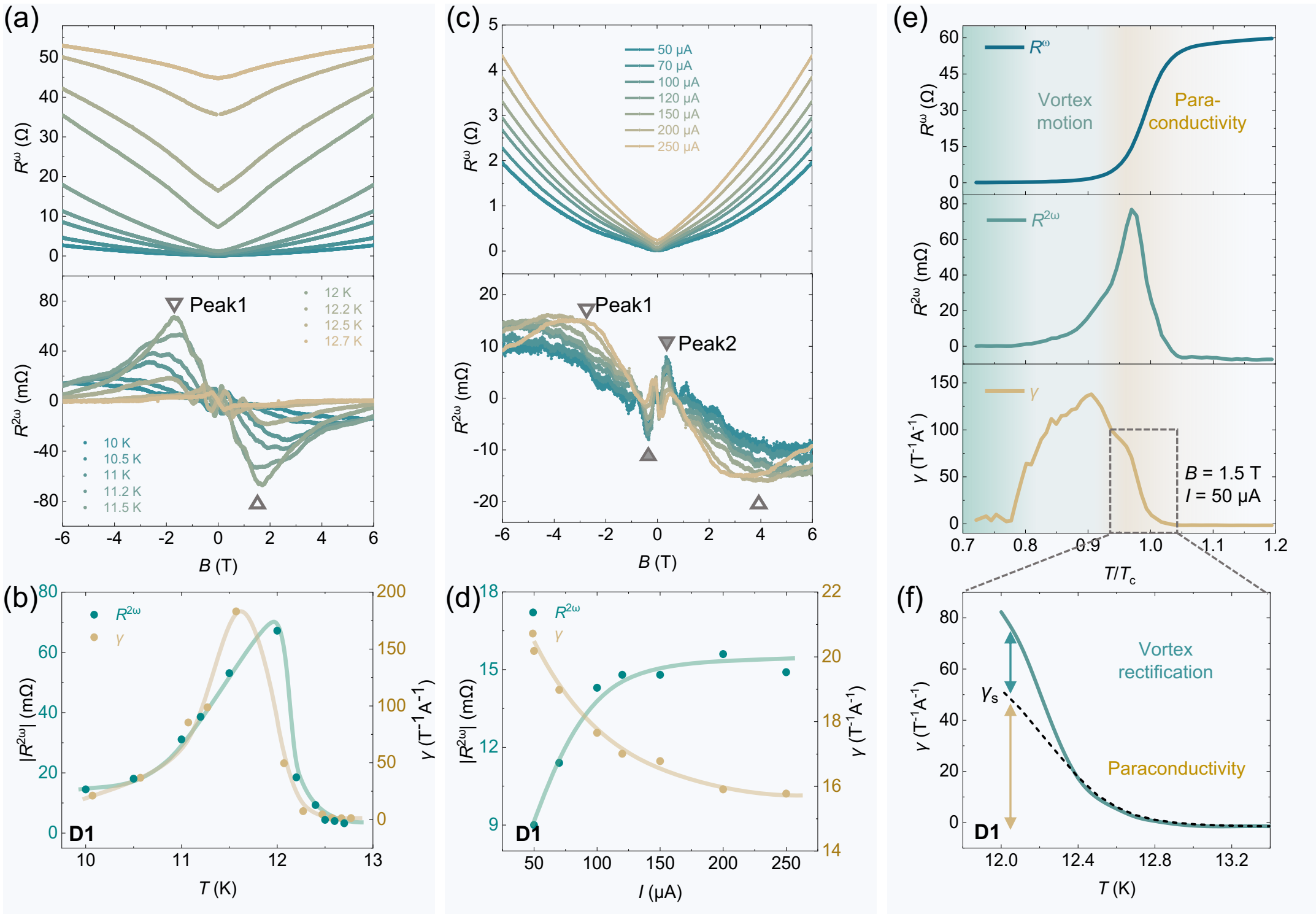

FIG. 4. Temperature and current dependence of the nonreciprocal transport. (a) Magnetic field dependence of $R^{\omega}$ and $R^{2\omega}$ at temperatures ranging from 10 to 12.7 K. (b) Extracted peak values of $|R^{2\omega}|$ and the corresponding $\gamma$ as a function of temperature. (c) Magnetic field dependence of $R^{\omega}$ and $R^{2\omega}$ at $T$ = 10 K for current ranging from 50 to 250 μA. (d) Extracted peak values of $|R^{2\omega}|$ (Peak1) and the corresponding $\gamma$ as a function of current amplitude. (e) Temperature dependence of $R^{\omega}$, $R^{2\omega}$, and $\gamma$ under $B$ = 1.5 T, $I$ = 50 μA at the vicinity of superconducting transition. (f) Temperature dependence of $\gamma$ near $T_c$, with fitting curves to separate contributions from paraconductivity and vortex dynamics.

## CONCLUSIONS

In this study, we demonstrate nonreciprocal charge transport in a strained FTS nanosheet. The formation of the cusp-like kink boundaries generates a spatially nonuniform strain profile with asymmetric distribution across the boundary, giving rise to an in-plane polarity that is absent in flat FTS nanosheets. We show that this asymmetric strain-induced polarity plays a key role in enabling nonreciprocal charge transport, which is strongly enhanced near the superconducting transition. Systematic temperature-, magnetic field- and current-dependent measurements reveal that the enhanced response originates from the combined contributions of vortex rectification and superconducting fluctuation paraconductivity. These results establish strain engineering via kink boundary formation as an effective route to induce and tune nonreciprocal charge transport in otherwise centrosymmetric superconductors, thereby expanding the material and design space for superconducting rectification devices.

## EXPERIMENTAL DETAILS

CVD growth of $FeTe_{0.7}Se_{0.3}$ with kink boundaries

In the experimental setup, 500 mg of Tellurium powder and 200 mg of Selenium powders were positioned upstream as the Te/Se sources. Simultaneously, a mixture of 10 mg powders containing $Fe_2O_3$ and $FeCl_2$ in a mass ratio of 5:1 was placed in an alumina boat at the center of the furnace. The 280 nm $SiO_2$/Si with the polished side facing downward, were positioned downstream in the tube. 80 sccm of Ar and 6 sccm of $H_2$ were used as carrier gases, being introduced into the 1-inch quartz tube under ambient pressure. The furnace was then heated to 520°C and maintained at this temperature for 5 minutes, and then rapidly cooled to room temperature using a fan to synthesize the $FeTe_{0.7}Se_{0.3}$ nanosheets with kink boundaries.

Transport measurement

The low-temperature transport measurements were conducted in an Oxford TeslatronPT cryostat under a magnetic field of 8 T. The temperature-dependent resistance curves were measured using a Keithley 6221 triggered with a Keithley 2182, at a frequency of 21 Hz. The *V-I* curves were obtained using a Keithley 2400 source meter to apply the d.c. current and a Keithley 2000 multimeter to measure the voltage. The $R^{\omega}$ and $R^{2\omega}$ were measured simultaneously using a Keithley 6221 source meter and two SR830 lock-in amplifiers with a 17.7 Hz sine wave AC current. Based on the principle, we symmetrized the $R^{\omega}$ raw data and antisymmetrized the $R^{2\omega}$ raw data.

**ACKNOWLEDGMENTS**

X.R.W. acknowledges support from the Academic Research Fund Tier 2 (Grant No. MOE-T2EP50210-0006 and MOE-T2EP50220-0016), Tier 3 (Grant No. MOE2018-T3-1-002) and Tier 3 (Grant No. MOE-MOET32023-0003) from Singapore Ministry of Education, and Agency for Science, Technology and Research (A*STAR) under its AME IRG grant (Project No. A20E5c0094). Z.L. acknowledges the support from Singapore National Research Foundation (NRF-CRP22-2019-0007, NRF-CRP26-2021-0004, NRF-F-CRP-2024-0008 and NRF-CRP31-0001), Singapore Ministry of Education Tier 3 Programme “Quantum Geometric Advantage” AcRF Tier 3 (MOE-MOET32023-0003), and Singapore Ministry of Education under its Research Centre of Excellence award to the Institute for Functional Intelligent Materials (EDUNC-33-18-279-V12). Q.W. acknowledges the support from National Natural Science Foundation of China (Grant No. 52271161).

**DATA AVAILABILITY**

There are no publicly available research data or software supporting this manuscript. Requests for further information or data should be sent to the authors.

References

[1] Y. Tokura and N. Nagaosa, Nonreciprocal responses from non-centrosymmetric quantum materials, Nat. Commun. **9**, 3740 (2018).

[2] M. Castellani, O. Medeiros, A. Buzzi, R. A. Foster, M. Colangelo, and K. K. Berggren, A superconducting full-wave bridge rectifier, Nat. Electron. **8**, 417 (2025).

[3] J. Ingla-Aynés, Y. Hou, S. Wang, E.-D. Chu, O. A. Mukhanov, P. Wei, and J. S. Moodera, Efficient superconducting diodes and rectifiers for quantum circuitry, Nat. Electron. **8**, 411 (2025).

[4] X. F. Lu, C.-P. Zhang, N. Wang, D. Zhao, X. Zhou, W. Gao, X. H. Chen, K. T. Law, and K. P. Loh, Nonlinear transport and radio frequency rectification in BiTeBr at room temperature, Nat. Commun. **15**, 245 (2024).

[5] N. Wang et al., Quantum-metric-induced nonlinear transport in a topological antiferromagnet, Nature **621**, 487 (2023).

[6] D. Nakamura, M.-K. Lee, K. Karube, M. Mochizuki, N. Nagaosa, Y. Tokura, and Y. Taguchi,

Nonreciprocal transport in a room-temperature chiral magnet, Sci. Adv. **11**, eadw8023 (2025).

[7] A. Dyrdał, J. Barnaś, and A. Fert, Spin-Momentum-Locking Inhomogeneities as a Source of Bilinear Magnetoresistance in Topological Insulators, Phys. Rev. Lett. **124**, 046802 (2020).

[8] S. Hoshino, R. Wakatsuki, K. Hamamoto, and N. Nagaosa, Nonreciprocal charge transport in two-dimensional noncentrosymmetric superconductors, Phys. Rev. B **98**, 054510 (2018).

[9] R. Wakatsuki, Y. Saito, S. Hoshino, Y. M. Itahashi, T. Ideue, M. Ezawa, Y. Iwasa, and N. Nagaosa, Nonreciprocal charge transport in noncentrosymmetric superconductors, Sci. Adv. **3**, 054510 (2017).

[10] J. Lustikova, Y. Shiomi, N. Yokoi, N. Kabeya, N. Kimura, K. Ienaga, S. Kaneko, S. Okuma, S. Takahashi, and E. Saitoh, Vortex rectenna powered by environmental fluctuations, Nat. Commun. **9**, 4922 (2018).

[11] E. Zhang et al., Nonreciprocal superconducting $NbSe_2$ antenna, Nat. Commun. **11**, 5634 (2020).

[12] F. Ando, Y. Miyasaka, T. Li, J. Ishizuka, T. Arakawa, Y. Shiota, T. Moriyama, Y. Yanase, and T. Ono, Observation of superconducting diode effect, Nature **584**, 373 (2020).

[13] Z. Yang, S. Li, S. Peng, X. Wang, L. Wu, R. He, Z. Wang, Y. Cao, and X. Renshaw Wang, Superconducting diode effect in gradiently strained $Nb_{0.5}Ti_{0.5}N$ films, Appl. Phys. Rev. **12**, 031411 (2025).

[14] S. Li, Z. Yang, X. Wang, K. P. Loh, and X. R. Wang, Nonreciprocal electrical transport in emerging noncentrosymmetric systems: from hidden symmetry to functional devices, Reports on Progress in Physics **89**, 066501 (2026).

[15] G. L. J. A. Rikken and E. Raupach, Observation of magneto-chiral dichroism, Nature **390**, 493 (1997).

[16] G. L. J. A. Rikken, J. Fölling, and P. Wyder, Electrical Magnetochiral Anisotropy, Phys. Rev. Lett. **87**, 236602 (2001).

[17] S. Li et al., Phase diagram of superconducting vortex ratchet motion in a superlattice with noncentrosymmetry, Phys. Rev. B **106**, 224513 (2022).

[18] T. Sato, H. Goto, and H. M. Yamamoto, Sturdy spin-momentum locking in a chiral organic superconductor, Phys. Rev. Res. **7**, 023056 (2025).

[19] H. Narita et al., Field-free superconducting diode effect in noncentrosymmetric superconductor/ferromagnet multilayers, Nat. Nanotechnol. **17**, 823 (2022).

[20] L. Du, T. Hasan, A. Castellanos-Gomez, G. Bin Liu, Y. Yao, C. N. Lau, and Z. Sun, Engineering symmetry breaking in 2D layered materials, Nature Reviews Physics **3**, 193 (2021).

[21] S. Yang, Y. Chen, and C. Jiang, Strain engineering of two-dimensional materials: Methods, properties, and applications, InfoMat **3**, 397 (2021).

[22] Y. Chen et al., Mobility Enhancement of Strained $MoS_2$ Transistor on Flat Substrate, ACS Nano **17**, 14954 (2023).

[23] Y. Dong, M.-M. Yang, M. Yoshii, S. Matsuoka, S. Kitamura, T. Hasegawa, N. Ogawa, T. Morimoto, T. Ideue, and Y. Iwasa, Giant bulk piezophotovoltaic effect in 3R-$MoS_2$, Nat. Nanotechnol. **18**, 36 (2023).

[24] J. Son, K.-H. Kim, Y. H. Ahn, H.-W. Lee, and J. Lee, Strain Engineering of the Berry Curvature Dipole and Valley Magnetization in Monolayer $MoS_2$, Phys. Rev. Lett. **123**, 036806 (2019).

[25] Z. Zhang et al., Near-100% spontaneous rolling up of polar van der Waals materials, Nat. Mater. **24**, 1716 (2025).

[26] A. P. Rooney, Z. Li, W. Zhao, A. Gholinia, A. Kozikov, G. Auton, F. Ding, R. V. Gorbachev, R.

J. Young, and S. J. Haigh, Anomalous twin boundaries in two dimensional materials, Nat. Commun. **9**, 1 (2018).

[27] F. Qin, W. Shi, T. Ideue, M. Yoshida, A. Zak, R. Tenne, T. Kikitsu, D. Inoue, D. Hashizume, and Y. Iwasa, Superconductivity in a chiral nanotube, Nat. Commun. **8**, 14465 (2017).

[28] Y. M. Itahashi, T. Ideue, S. Hoshino, C. Goto, H. Namiki, T. Sasagawa, and Y. Iwasa, Giant second harmonic transport under time-reversal symmetry in a trigonal superconductor, Nat. Commun. **13**, 1659 (2022).

[29] T. Ideue, S. Koshikawa, H. Namiki, T. Sasagawa, and Y. Iwasa, Giant nonreciprocal magnetotransport in bulk trigonal superconductor $PbTaSe_2$, Phys. Rev. Res. **2**, 042046 (2020).

[30] Y. Imai, Y. Sawada, F. Nabeshima, and A. Maeda, Suppression of phase separation and giant enhancement of superconducting transition temperature in $FeSe_{1-x}Te_x$ thin films, Proceedings of the National Academy of Sciences **112**, 1937 (2015).

[31] Z. Ge, Q. Zou, H. Zhang, C. Yan, D. Agterberg, M. Weinert, and L. Li, Superconductivity on Edge: Evidence of a One-Dimensional Superconducting Channel at the Edges of Single-Layer FeTeSe Antiferromagnetic Nanoribbons, ACS Nano **14**, 6539 (2020).

[32] D. Hu et al., Chemical Vapor Deposition of Superconducting $FeTe_{1-x}Se_x$ Nanosheets, Nano Lett. **21**, 5338 (2021).

[33] S. Li et al., Field-Free Superconducting Diode Effect and Magnetochiral Anisotropy in $FeTe_{0.7}Se_{0.3}$ Junctions with the Inherent Asymmetric Barrier, ACS Nano **18**, 31076 (2024).

[34] M.-M. Yang, D. J. Kim, and M. Alexe, Flexo-photovoltaic effect, Science (1979). **360**, 904 (2018).

[35] J. Jiang, Z. Chen, Y. Hu, Y. Xiang, L. Zhang, Y. Wang, G.-C. Wang, and J. Shi, Flexo-photovoltaic effect in MoS2, Nat. Nanotechnol. **16**, 894 (2021).

[36] See Supplemental Material at [URL] for transport measurements under different current and magnetic-field orientations performed on multiple devices.

[37] T. Yokouchi, N. Kanazawa, A. Kikkawa, D. Morikawa, K. Shibata, T. Arima, Y. Taguchi, F. Kagawa, and Y. Tokura, Electrical magnetochiral effect induced by chiral spin fluctuations, Nat. Commun. **8**, 1 (2017).

[38] C. Guo et al., Switchable chiral transport in charge-ordered kagome metal $CsV_3Sb_5$, Nature **611**, 461 (2022).

[39] Y. M. Itahashi, T. Ideue, Y. Saito, S. Shimizu, T. Ouchi, T. Nojima, and Y. Iwasa, Nonreciprocal transport in gate-induced polar superconductor $SrTiO_3$, Sci. Adv. **6**, eaay9120 (2020).

[40] K. Yasuda, H. Yasuda, T. Liang, R. Yoshimi, A. Tsukazaki, K. S. Takahashi, N. Nagaosa, M. Kawasaki, and Y. Tokura, Nonreciprocal charge transport at topological insulator/superconductor interface, Nat. Commun. **10**, 2734 (2019).

[41] Y. M. Itahashi, Y. Saito, T. Ideue, T. Nojima, and Y. Iwasa, Quantum and classical ratchet motions of vortices in a two-dimensional trigonal superconductor, Phys. Rev. Res. **2**, 023127 (2020).

[42] C. C. de Souza Silva, J. Van de Vondel, M. Morelle, and V. V. Moshchalkov, Controlled multiple reversals of a ratchet effect, Nature **440**, 651 (2006).

[43] Y.-L. Wang, A. Glatz, G. J. Kimmel, I. S. Aranson, L. R. Thoutam, Z.-L. Xiao, G. R. Berdiyorov, F. M. Peeters, G. W. Crabtree, and W.-K. Kwok, Parallel magnetic field suppresses dissipation in superconducting nanostrips, Proceedings of the National Academy of Sciences **114**, E10274 (2017).

[44] C. C. de Souza Silva, A. V. Silhanek, J. Van de Vondel, W. Gillijns, V. Metlushko, B. Ilic, and V.

V. Moshchalkov, Dipole-Induced Vortex Ratchets in Superconducting Films with Arrays of Micromagnets, Phys. Rev. Lett. **98**, 117005 (2007).

[45] W. Gillijns, A. V. Silhanek, V. V. Moshchalkov, C. J. O. Reichhardt, and C. Reichhardt, Origin of Reversed Vortex Ratchet Motion, Phys. Rev. Lett. **99**, 247002 (2007).